# The Monitor project: tracking the evolution of low mass and pre-main sequence stars

Suzanne Aigrain[1], Jonathan Irwin[2], Leslie Hebb[3], Simon Hodgkin[4], Adam Miller[5], Estelle Moraux[6], Keivan Stassun[7]

[1]School of Physics, University of Exeter, UK
[2]Harvard-Smithsonian Center for Astrophysics, Cambridge, MA, USA
[3]School of Physics and Astronomy, University of St Andrews, UK
[4]Institute of Astronomy, University of Cambridge, UK
[5]Astronomy Department, University of California at Berkeley, CA, USA
[6]Laboratoire d'Astrophysique de Grenoble, France
[7]Department of Physics and Astronomy, Vanderbilt University, Nashville, TN, USA

**The Monitor project is a large-scale program of photometric and spectroscopic monitoring of young open clusters using telescopes at ESO and other observatories. Its primary goal is to detect and characterise new low-mass eclipsing binaries, and the first three detected systems are discussed here. We derive the masses and radii of the components of each system directly from the light and radial velocity curves, and compare them to the predictions of commonly used theoretical evolutionary models of low-mass stars.**

### Eclipsing binaries as calibrators of stellar evolution

Well-observed detached eclipsing binaries (EBs) are extremely valuable systems, because it is possible to derive accurate (to a few percent) model-independent estimates of the individual masses and radii and the temperature ratio directly from the light curve and radial velocity (RV) curves of each component. This is particularly true in open clusters, where the age and chemical composition of the stars is well-known, and even more so in young clusters and star forming regions, where 1) one probes a phase of rapid stellar evolution, as highlighted by Figure 1 in the mass-radius plane, and 2) low-mass systems, which are of particular interest because the physics and chemistry of their convective interiors and cool atmospheres are complex to model, are still relatively bright. Yet, pre-2004, there were no sub-solar pre-main sequence EBs known, and very few on the main sequence.

This was the principal motivation for starting the Monitor project[1], a systematic survey for low-mass EBs in nine young (less than 200 Myr), rich and relatively nearby star forming regions and open clusters. The survey proceeds in two phases, starting with intensive I-band photometric monitoring to detect candidate eclipsing systems, followed by multi-epoch spectroscopic observations to measure radial velocities, and hence derive component masses and determine if the systems belong to the clusters or are older field objects. A strong additional motivation was the possibility of detecting young transiting planets, which would give important insights into planet formation timescales and the initial conditions for giant planet evolution.

### Cluster photometry

The full target list and detail of the photometric observations are given in Aigrain et al. (2007). For this part of the project, we use 2 to 4m telescopes equipped with wide-field optical imagers. This enables us to efficiently monitor large fractions of each cluster, while ensuring both the photometric precision and the time sampling necessary to detect eclipses down to mid-M spectral types. Typically, each cluster must be observed for about 100 hours in total. A major component of the photometric observations were carried out as part of an ESO Large Programme[2] using the Wide Field Imager (WFI) on the 2.2m ESO/MPI telescope between June 2005 and May 2006.

All the Monitor photometric data are reduced automatically using a uniform procedure, described in detail in Irwin et al. (2007a). Briefly, we first carry out all the standard CCD data reduction steps and astrometric and photometric calibration using the Cambridge Astronomical Survey Unit (CASU) pipeline. We then perform simple aperture photometry to generate light curves, but with a number of precautions designed to maximise photometric precision and minimise correlated systematic effects, often termed red noise, which are known to have a major impact on the yield of planetary transit surveys (Pont et al. 2006), and by extension mixed eclipse and transit surveys such as Monitor. The resulting precision illustrated in Figure 2 for the WFI observations of NGC 2547. We achieve sub-percent relative photometry over four magnitudes or more from the saturation limit for all Monitor clusters. The noise level over a typical eclipse timescale of 2.5 hours varies between 1.5 and 3 mmag (depending on the instrument and observing conditions) for the brightest objects in each cluster.

---

[1] http://www.ast.cam.ac.uk/research/monitor/
[2] Advanced data products (reduced images and source catalogs) from the first run of this program (LP 175.C-0685) have recently been made available via the ESO archive.

For each cluster, we also take long V-band exposures which, combined with a stacked selection of the best I-band images, enables us to construct a deep colour-magnitude diagram (CMD). We identify likely members of the cluster as those lying close to an empirically defined cluster sequence on this CMD, allowing for the fact that cluster binaries are overluminous for their colour, and for the increased uncertainties towards the faint end. This results in candidate membership lists with a global estimated level of contamination by field stars varying between 30 and 80%, depending on whether the cluster lies in the Galactic plane.

The light curves of candidate members and then systematically searched for eclipses, using a combination of visual examination and an automatic procedure (Miller et al. submitted) involving a pre-filtering step to remove the significant spot-induced variability of our young, active targets where possible, followed by an automated box-shaped eclipse search algorithm. In this fashion, we have so far identified 48 high quality EB candidates, ~15% of which have depths compatible with a planetary companion.

**Spectroscopic follow-up**

Once EB candidates have been identified, multi-epoch medium- to high-resolution spectroscopy is needed to determine RV orbits for both components, and thus derive their masses and determine (by comparing the systemic velocity to that of the cluster) whether the detected systems are really cluster members or older field systems projected onto the cluster. The latter can still be of interest, as our CMD cut typically implies they are low-mass systems.

Initially, we opted for a 2-step, risk-minimising strategy, starting with medium resolution (R~10000) resolution spectroscopy on 4m-class telescopes, including EMMI on the NTT and ISIS on the William Herschel Telescope (WHT), to weed out obvious non-members and derive orbits for the systems with higher RV amplitudes and a more favourable flux contrast between primary and secondary. For the remaining systems, we then used higher resolution (R~20000 to 50000) observations on 8m-class telescopes, including FLAMES on the VLT and Phoenix on Gemini-South, to resolve the two components and/or detect lower amplitude RV variations. Since virtually all of the candidates we have followed-up so far with sufficient precision have turned out to be binaries, in future we will concentrate on VLT/FLAMES observations for all candidates, the multiplex capability of FLAMES allowing us to exploit optimally the spatial concentration of our candidates. In all cases, RVs are derived by cross-correlation with simultaneously observed RV standards or model spectra of appropriate spectral type.

**Three new low mass eclipsing binary systems**

The spectroscopic follow-up to date has enabled us to derive partial orbits for over a dozen systems in 6 different cluster fields, and full solutions for 3 previously unknown systems. JW380 (Irwin et al. 2007b) is a member of the ~1 Myr old Orion Nebula Cluster (ONC). The other two are likely field systems (as indicated by their systemic velocities and confirmed by the relatively compact radii derived for their components) projected onto the fields of NGC 2362 and NGC 2547 respectively, and are the subject of publications in preparation. For simplicity, they are referred to from now on as EB2 and EB3.

The RV data for each system were modelled assuming Keplerian orbits (see Figure 3). The period and phase were fixed at the values determined from the light curve. The best-fit eccentricities are zero within the uncertainties, and the fits shown therefore assume circular orbits. These fits were used to derive mass ratios and a minimum total mass (the RV data alone does not yield the inclination of the orbit). The I-band light curves were modelled separately (see Figure 4) using the JKTEBOP[3] code after manual removal of most of the out-of-eclipse variability (assigned to surface spots) to derive inclinations and individual radii. Estimates of the main system parameters are summarised in Table 1, but the reader is referred to the relevant individual publications for more details.

When these systems are placed on a mass-radius diagram (Figure 1) and compared with other known systems and theoretical isochrones, it becomes apparent that each of probes a distinct, but so-far ill-constrained mass range. JW380 is one of only a handful of sub-solar pre-main sequence systems, and one of the two youngest (both are in the ONC). At $M_1 = 0.26\ M_\odot$ and $M_2 = 0.15\ M_\odot$, it is also one of the lowest-mass EBs known. The 2 and 3 Myr NextGen isochrones of the Lyon group are in reasonable agreement with the measured parameters of its components (although the canonical age for the ONC is 1 Myr, a significant age spread is thought to be present in the region, so an age of 2 or 3 Myr is not inconceivable for JW 380). More photometric and spectroscopic observations of this system are planned to better constrain the component masses, attempt to derive their temperatures, and refine the radii by sampling the out-of-eclipse variations more tightly.

---

[3] http://www.astro.keele.ac.uk/~jkt/codes/jktebop.html

EB2 and EB3, with component masses of 0.29 $M_\odot$ and 0.17 $M_\odot$, and 0.62 $M_\odot$ and 0.25 $M_\odot$ respectively, fill gaps in the existing empirical main-sequence relation. Given our preliminary radius estimates, the main-sequence isochrone fits the secondaries of both field systems well, but the primaries appear significantly larger than expected. In other words, for a given luminosity, these primaries appear too cool for their mass and age. This phenomenon has already been seen in other systems, and the young eclipsing binary brown dwarf 2MASS J05352184-0546085 (Stassun et al. 2006), whose primary is cooler than the secondary, may be an extreme example. It is not yet clear what causes this effect, but it could be a consequence of the high magnetic fields expected in the rapidly rotating components of close binaries (Chabrier et al. 2007). These might inhibit convection in the interior, and hence heat transport to the surface of the star, and do so more strongly the more massive of the two stars. Additional photometric observations are planned for EB2 and EB3 to refine the radius determinations.

**Additional science from Monitor**

High cadence, high precision time-series photometry of large samples of young stars of known age allows for a wealth of ancillary science going far beyond the sole search for eclipsing binary systems. In particular, we have so far used Monitor data in 4 of our target clusters (M34, NGC 2516, NGC 2547 and NGC 2362) to identify over 1500 new photometric rotation periods for low-mass stars. Combined with existing data from the literature in other clusters, these now constitute a well-populated age sequence, from the birth-line to the start of main-sequence evolution, with detailed implications for models of the angular momentum evolution of low-mass stars, which is intrinsically linked to the evolution of their internal structure and, in the initial phases, coupling with their circumstellar disks (for a review of the Monitor rotation results, see Bouvier 2007 and references therein).

Because of the favourable radius ratio for low-mass primaries, it is also possible that some of our remaining candidate eclipsing systems have secondaries in the planetary mass range. If so, they would provide the first radius measurement(s) for young, close-in giant planet(s). Already, we have used observations of NGC 2362 to derive statistically significant upper limits on the incidence of such planets at 5 Myr (Miller et al. in prep.).

**Future Prospects**

The process of following-up our candidate list is ongoing, with ongoing FLAMES of the ONC and 3 other clusters. We are also using these observations to carry out systematic spectroscopic binary searches in the clusters, in order to measure (near-)primordial close binary fractions and mass ratio distributions for low-mass stars, and to investigate the connection between rotation, accretion and binarity.

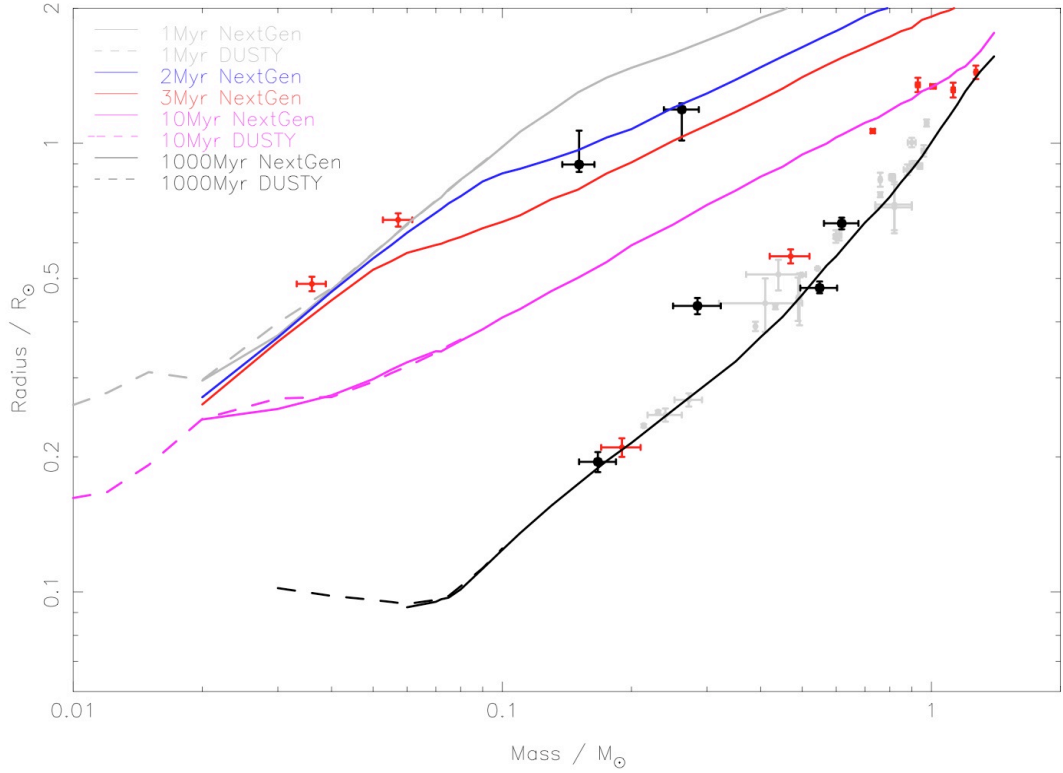

**Figure 1:** Age-mass-radius relation for low-mass stars and brown dwarfs in eclipsing binaries. Grey and red symbols represent field and pre-main sequence systems from the literature, respectively (see Irwin et al. 2007b for a full reference list). The three new systems detected by Monitor are shown in black. The solid and dashed lines show the NextGen and DUSTY isochrones of the Lyon group for 1, 2, 3, 10 and 1000 Myr (grey, blue, red, pink and black respectively). The black isochrone essentially corresponds to the main sequence relation).

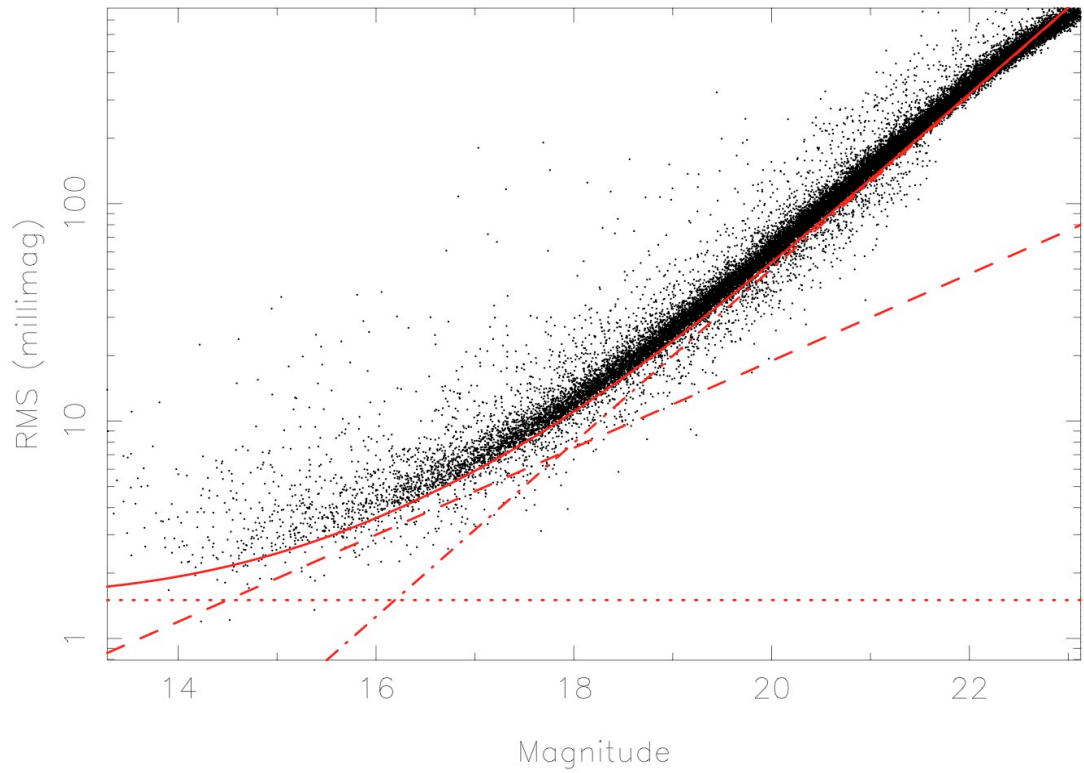

***Figure 2:*** *Photometric precision versus I-band magnitude for the WFI observations of NGC2547 (100h in service mode spanning 8 months from October 2005 to May 2006). The dashed, dotted, dash-dot and solid lines represent the source photon noise, sky photon noise, a 1.5mmag constant added to account for systematics, and the final expected precision (quadrature sum of the other three lines) respectively.*

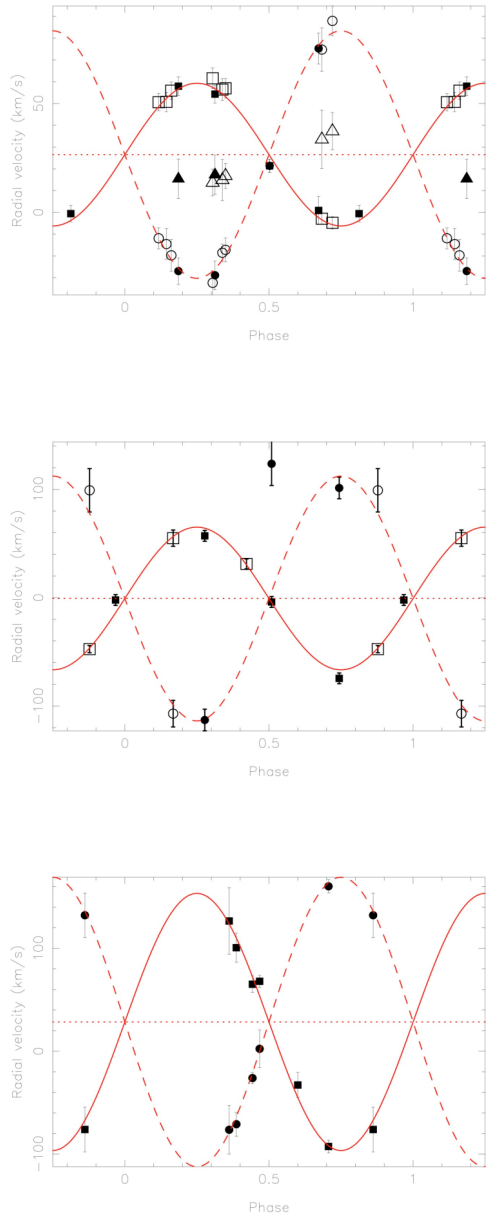

*Figure 3:* RV data and best-fit model for the 3 new Monitor EBs. Field symbols represent VLT/FLAMES data for JW 380 (left/top as appropriate) and NTT/EMMI data fro EB2 (centre) and EB3 (right/bottom as appropriate), hollow symbols data Gemini/Phoenix and ISI/WHT data respectively. Velocities for the primary, secondary and (if present) tertiary of the system are shown as squares, circles and triangles respectively. The solid, dashed and dotted red lines show the best fit orbits for the primary and secondary and the systemic velocity respectively.

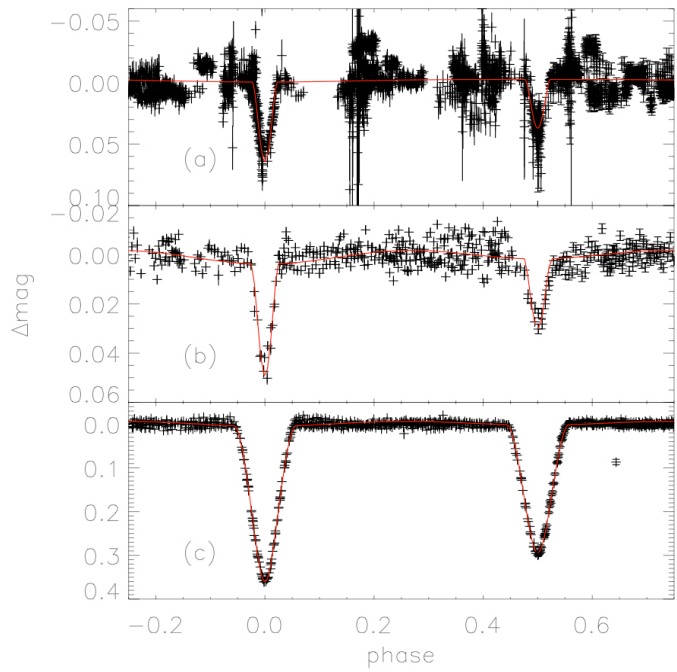

*Figure 4:* I-band relative light curve and JKTEBOP best-fit model for each system: (a) JW 380 (ONC) I=13.82, data from the Wide Field Camera on the 2.4m Isaac Newton telescope; (b) EB in the field of NGC 2362, I=15.6 , data from MosaicII on the Blanco 4m telescope at Cerro Tololo Interamerican Observatory; (c) EB in the field of NGC 2547, I=15.5 , data from WFI on the ESO/MPI 2.2m telescope.

| Object | $M_1$ ($M_\odot$) | $M_2$ ($M_\odot$) | $R_1$ ($R_\odot$) | $R_2$ ($R_\odot$) | P (d) |
|---|---|---|---|---|---|
| JW 380 | 0.262±0.025 | 0.151±0.013 | $1.189^{+0.039}_{-0.175}$ | $0.897^{+0.170}_{-0.034}$ | 5.29918 |
| EB2 | 0.285±0.038 | 0.167±0.017 | 0.434±0.018 | 0.195±0.010 | 0.71362 |
| EB3 | 0.618±0.018 | 0.549±0.054 | 0.663±0.020 | 0.476±0.016 | 0.58530 |

**Table 1:** Derived primary mass ($M_1$), secondary mass ($M_2$), primary radius ($R_1$), secondary radius ($R_2$) and period (P) for each system.